\documentclass[aps,prl,showpacs,twocolumn,floats,superscriptaddress,10pt]{revtex4-1}
\usepackage{graphicx}
\usepackage{dcolumn}
\usepackage{bm}
\usepackage{color}
\usepackage{amssymb}
\begin{document}
\title{Kondo Force in Shuttling Devices:
Dynamical Probe for a Kondo Cloud}
\author{M.N. Kiselev}
\affiliation{The Abdus Salam International Centre for Theoretical Physics,
Strada Costiera 11, I-34151 Trieste, Italy}
\author{K.A. Kikoin}
\affiliation{School of Physics and Astronomy, Tel-Aviv University, Tel-Aviv
69978, Israel}
\author{L.Y. Gorelik}
\affiliation{Chalmers University of Technology, Department of Applied Physics, SE-412 96 G\"oteborg, Sweden}
\author{R.I. Shekhter}
\affiliation{University of  Gothenburg, Department of Physics, SE-412 96 G\"oteborg, Sweden}
\date{\today}
\pacs{73.23.-b,
  72.10.Fk,
  73.23.Hk,
  85.85.+j}
\begin{abstract}
We consider electromechanical properties of a single-electronic device consisting of
movable quantum dot attached to a vibrating cantilever, forming a tunnel contact with a non-movable
source electrode. We show that the resonance Kondo tunneling of electrons amplify exponentially the
strength of nanoelectromechanical (NEM) coupling in such device and makes the latter to be insensitive to mesoscopic fluctuations of electronic levels in a nano-dot.
It is also shown that the study of Kondo-NEM phenomenon provides an additional (as
compared with a standard conductance measurements
in a non-mechanical device) information on retardation effects in formation of
many-particle cloud accompanied the Kondo tunneling. A possibility
for superhigh tunability of mechanical dissipation as well as supersensitive
detection of mechanical displacement is demonstrated.
\end{abstract}
\maketitle

Recent progress in fabrication of nanoelectromechanical systems (NEMS)
\color{black} based on suspended carbon nanotubes \cite{Hut09} 
as well as on suspended Si
\cite{Azuma07}  and SiN \cite{Weig12} \color{black} nanowires 
vibrating at radio frequencies (RF) resulted in rapidly growing amount of theoretical works
\cite{Sheht05}-\cite{Shekht10} addressing the issues of interplay between spin/charge transport
and nano-mechanics \cite{kis06}. The observation of Coulomb Blockade \cite{Azuma07} in NEMS
opened a possibility to consider the influence of strongly correlated and resonance effects
on a behaviour of nano-oscillators.

Usually, NEM regime implies strong coupling between the electronic and
mechanical degrees of freedom.
The coupling  is provided by two main mechanisms.
Motion of the movable dot (shuttle) between two metallic banks results in the time dependent
tunneling amplitudes. On the other hand,  
the electron charge transport between the banks in presence of magnetic field
results in appearance of a Lorentz/Laplace force acting on the shuttle,
which should also be taken into account \cite{Sheht05,Sheht09}.
The aggregate dynamics of a shuttle is that of a periodic
oscillator with decrement or increment and an electron tunneling
(cotunneling) parametrically dependent on this slow classical
motion. In some sense the problem may be treated as a tunneling
through anharmonic vibronic system. In many cases, e.g. in
shuttling devices including bending carbon nanotubes
\cite{Sheht09}, the vibronic language completely describes the
physical situation.
\begin{center} 
\begin{figure}[b]
\vspace*{-10mm}
\includegraphics[angle=0,width=\columnwidth]{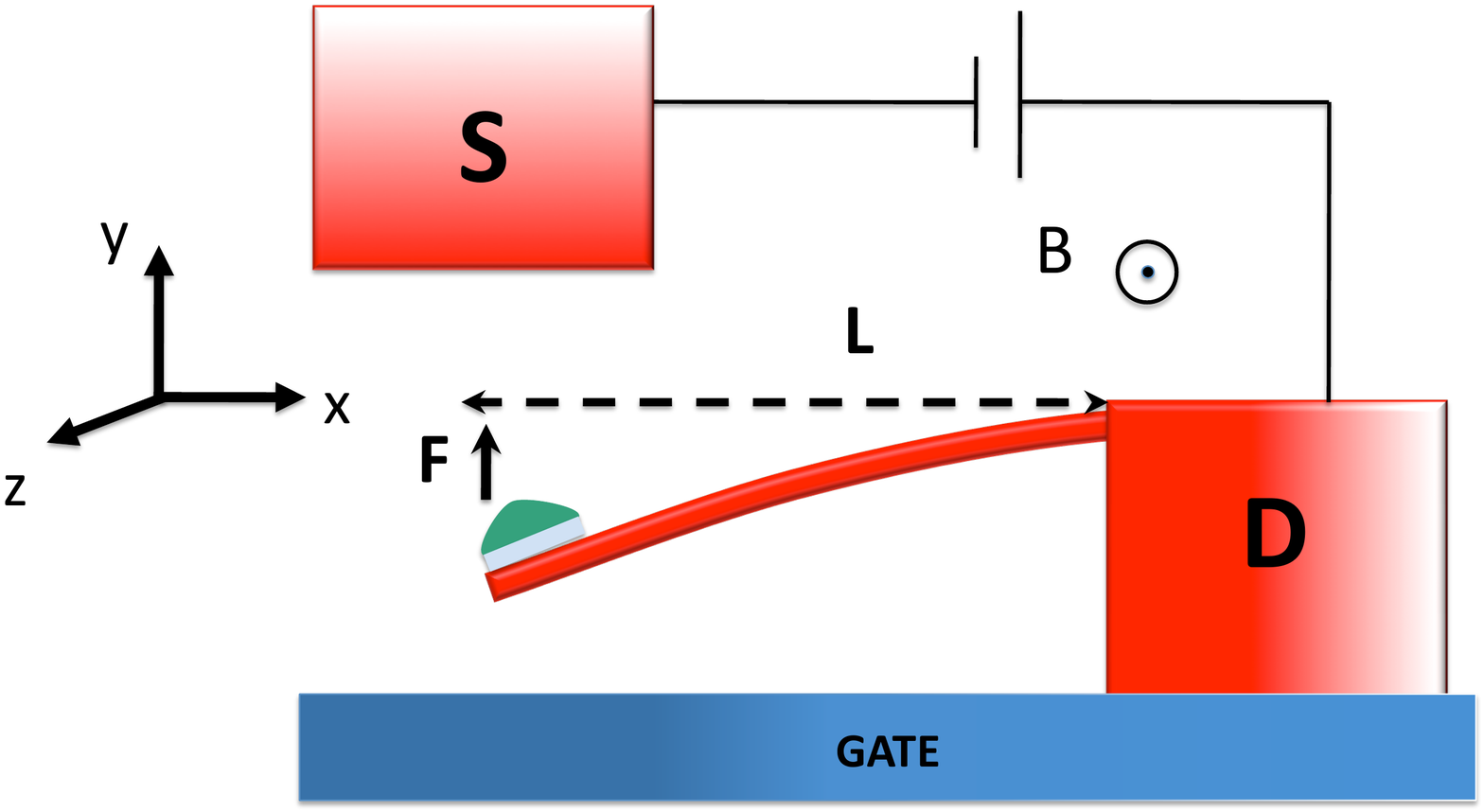}
\vspace*{-15mm}
\caption{(Color online) Shuttle with a cantilever. The nanoisland (green) is separated from the wire
(drain) by a tunnel barrier (light blue) of constant width. The width of the barrier to the source 
changes during the cycle. Shuttle oscillations are stimulated by initial conditions, e.g. bending the cantilever 
by the gate voltage at time $t=0$.} \label{f.1}
\end{figure}
\end{center}
\vspace*{-7mm}
\par
NEM coupling like other nanometer length scale phenomena
is strongly affected by mesoscopic fluctuations. Spatial quantization of electronic motion in a quantum dot makes electro-mechanical transduction to be sample sensitive phenomenon \cite{Azuma07}. An exception 
to this rule is electromechanical coupling due to the many-body Kondo tunneling.
Indeed in this case the charge transfer is controlled by the 
singularity of the tunneling density of states at the energy
pinned to the Fermi level of the injector and thus protected against mesoscopic fluctuations.
This fact in combination with another generic feature of Kondo phenomenon - its super-sensitivity to a strength of the tunneling coupling (and therefore its super-sensitivity to the mechanical displacement of a quantum dot) - makes Kondo NEM coupling to be a phenomenon promising for practical applications.

An example of such device 
is schematically shown in Fig. \ref{f.1}.
A nanoisland is mounted on the metallic cantilever, which may
vibrate under an external force. The contact between the source and
drain electrodes is a combination of time-dependent tunneling bridge 
between the source $S$ and the island and a metallic bridge formed by vibrating cantilever
connecting the island with the drain $D$ (see Ref.
\onlinecite{Azuma07} for experimental realization).

We consider the configuration where a cantilever is
displaces in $y$ direction, $\vec u =(0,y,0)$ in a magnetic field
$\vec B = (0,0,B)$. In this case the
Laplace force $\vec F$ acts on the
cantilever in the same direction $y$,  $\vec F_L = L\cdot\vec I\times \vec B =(0,F,0)$. 
Here $L$ is the length of the cantilever. 
Besides, small electromotive force $\vec F_{\rm emf} = (f,0,0)$
acts on the electrons in the cantilever. 
In the limit of strong Coulomb blockade in the nanoisland, the
Kondo screening accompanies the tunneling 'source-island-cantilever', and a
unique possibility arises to study the contribution of purely
quantum many-particle Kondo effect on the classical oscillation of
a shuttle (cantilever + island). The study of a "Kondo force'' in
shuttling is the main subject of this paper.

We study two coupled
subsystems: a tunneling contact 'source-moving island-moving cantilever' treated
as a purely quantum system in a framework of the Anderson -- Kondo
model, and a macroscopic wire with attached island oscillating
under external constraining force. We work in the Kondo limit, where the nanoisland is
represented by its spin $\vec S$, so that internal degrees of
freedom are the spin-flip processes. The source-drain transport is
a combination of quantum tunneling 'source - island - cantilever' and ohmic
transport 'island-drain'. In this "time-dependent Schrieffer --
Wolff" limit (see below) the Hamiltonian of quantum subsystem is
\begin{eqnarray}
H=H_{\rm lead} + H_{\rm ex} +\delta H\nonumber
\end{eqnarray}
\begin{eqnarray}
H_{\rm lead} = \sum_{\alpha=l,r}\sum_{k\sigma}
\xi_kc^\dag_{\alpha k\sigma}
c^{}_{\alpha k\sigma},\;\;\;\;H_{\rm ex} = \sum_{\alpha\alpha'}J_{\alpha\alpha'}\vec
s_{\alpha\alpha'}\cdot \vec S 
\nonumber
\end{eqnarray}
\vspace*{-5mm}
\begin{eqnarray}\label{2.1}
\delta H = \frac{eV_{\rm bias}}{2}(N_l - N_r) 
\end{eqnarray}
Here the indices $l,r$ stand for the electronic states in the 
source and cantilever, respectively, 
$\xi_k = \varepsilon_k -\mu$ are the excitation energies of lead electrons,
$N_\alpha=\sum_{k\sigma} n_{\alpha k\sigma}$ are the corresponding electron density operators,
$ \vec s^{}_{\alpha\alpha'}= \frac{1}{2}
\sum_{kk'}c^\dag_{\alpha k\sigma}\vec\tau^{}_{\sigma\sigma'} c^{}_{\alpha' k'\sigma'},
 ~\vec S = \frac{1}{2}d^\dag_{m}\vec\sigma^{}_{mm'} d^{}_{m'}$
are the spin operators for the electrons in the leads and in the
nanoisland, respectively, $\vec \tau$ and $\vec \sigma$ are the vectors of Pauli
matrices acting on the states in the leads and dot. 
At small bias voltage $V_{\rm bias}$ the source and the cantilever  are
supposed to be in the adiabatically stationary state of thermal equilibrium. 
The parameters  are 
$J_{\alpha\alpha'}=4v^{*}_\alpha v^{}_{\alpha'}/E_c$,
where $v^{}_\alpha$ is the \color{black} tunneling amplitude \color{black} between the
nanoisland and the metallic lead $\alpha$, $E_c$ is the Coulomb
blockade energy. The \color{black} exchange couplings \color{black} $J^{}_{ll}$ and
$J^{}_{lr}$ are time-dependent due to the dependence of the
tunneling amplitude between the source and the moving nanoisland
on the island position $v^{}_l=v^{}_l[\vec u(t)]$. The time
dependence of this amplitude is a set of pulses corresponding to
electron injection from the metallic reservoir to the shuttle
periodically approaching the bank $S$.\cite{footnote} We 
confine our treatment with the simplest case of $S=1/2$ (odd
occupation of a nanoisland in the neutral state) and the single
channel tunneling between the nanoisland and  the leads.

The oscillations of cantilever with attached nanoisland are
determined by the classical Newton equations
\begin{equation}\label{2.4}
  \ddot{\vec u} + \frac{\omega_0}{Q_0} \dot{\vec u} + \omega^2_0\vec u =\frac{1}{m} \vec F.
\end{equation}
 where $\omega_0=\sqrt{k/m}$ is the oscillator frequency of free cantilever,
$Q_0$ is a quality factor of NEM device.

Our aim is to study the spin and charge transport by means of a
shuttle oscillating in accordance with Eq. (\ref{2.4}) in presence
of many-particle Kondo screening described by the Hamiltonian
(\ref{2.1}). The coupling between the classical and quantum
subsystem is realized via the parameters
$J_{ll}(\vec u),~ J_{lr}(\vec u),~ \vec F(\vec u),$
where the time dependence $\vec u(t)$ should be calculated self-consistently. \color{black}
Meanwhile, $J_{rr}$ does not depend on displacement  $\vec{u}$ (see Fig. 1).\color{black}

The cotunneling Hamiltonian may be rationalized by means of the Glazman-Raikh rotation, which in our
situation is time dependent:
\begin{equation}\label{2.5}
 \left(%
\begin{array}{c}
  c_{lk\sigma} \\
  c_{rk\sigma} \\
\end{array}%
\right)= \left(%
\begin{array}{cc}
  \cos \vartheta_t & -\sin\vartheta_t \\
  \sin\vartheta_t & \cos\vartheta_t \\
\end{array}%
\right)\left(%
\begin{array}{c}
  \psi_{1k\sigma} \\
  \psi_{2k\sigma} \\
\end{array}%
\right)\equiv U_t\left(%
\begin{array}{c}
  \psi_{1k\sigma} \\
  \psi_{2k\sigma} \\
\end{array}
\right)
\end{equation}
with $\tan \vartheta_t = |v_r/v_l(t)|$. Use of this transformation
for diagonalization of the Schr\"odinger operator ${\cal L} =
-i\hbar d/dt + H(t)$ results in generation of additional term
$H_B$ proportional to $-i\color{black}\hbar \color{black}U_t^{-1}\partial_t U$ in the transformed Hamiltonian (see, e.g., \cite{Aono04}).
\begin{eqnarray}\label{2.6}
H'= H^{}_{\rm lead} + H_{\rm B} + H_{\rm ex} + \delta H , 
\end{eqnarray}
\begin{eqnarray}\label{2.61}
 H_{\rm lead}&=&\sum_{a=1,2}\sum_{k\sigma} \xi_{k}\psi^\dag_{a k\sigma}
\psi^{}_{a k\sigma},\nonumber\\
 H_{\rm B}
 &=& i\color{black}\hbar\color{black}\frac{d\vartheta_t}{dt}\sum_{k\sigma}\left(
  \psi^\dag_{1k\sigma}\psi^{}_{2k\sigma}
  - \psi^\dag_{2k\sigma}\psi^{}_{1k\sigma}\right),\nonumber\\
  H_{\rm ex} &=&\frac{J}{\color{black}4\color{black}}\sum_{kk',\sigma\sigma',m,m'}%
  \psi^\dag_{1k\sigma} \vec\tau_{\sigma\sigma'}
  \psi_{1k'\sigma'}
d^\dag_{m}\vec \sigma_{m m'} d^{}_{m'},
\end{eqnarray}
$\delta H$$=$$\frac{\displaystyle eV_{\rm bias}}{\displaystyle 2}$$\big[$$($$N_2$$-$$N_1$$)$$\cos$$2$$\vartheta_t$$+$$\displaystyle \sum_{k\sigma}$$($$\psi^\dag_{1k\sigma}\psi^{}_{2k\sigma}$$+$$h.c.$$)$$\sin$$2$$\vartheta_t\big].$
The term $H_B$ may be treated as an additional gauge potential in
the lead Hamiltonian describing a \color{black} Berry-like \color{black} phase\cite{Aono04}
generated by shuttle motion \cite{com1}.  Only the even partial wave $\psi_1$
survives in the cotunneling term $H_{\rm ex}$ with the
time-dependent effective indirect exchange coupling $J (t)=
J_{ll}(t)+J_{\color{black}rr\color{black}}$ (see, e.g., \cite{PuGla04}). This time dependence
may be parametrized \cite{kis06} in assumption that
the source-island tunneling amplitude is an exponential function of a
distance $y$ between the source and the moving nanoisland, while
the tunneling nanoisland-cantilever is constant:
$v_r=v_0,~~ v_l = v_1\exp[y(t)/\lambda].$
The spatial coordinates are counted off the equilibrium position
of cantilever, so that $v_1$ is exponentially small, $v_1/v_0 \sim
\exp(-y_0/\lambda)$. Here $\lambda$ is the confinement radius (tunnel length) of
the electron wave function within the island, $y_0$ is the
distance between the source and the island at equilibrium.

We suppose that the shuttling mode is slow enough and the electron
transport is adiabatic, i.e., the \color{black} exchange couplings \color{black} $J_{ll}, J_{lr}$
depend parametrically on time via the displacement coordinates
$\vec u(t)$. Then to find the tunneling current one may trace the
time dependence of local occupations  of the left and right banks
(source and nanoisland) near the point of tunneling contact. 
The current operator is
\begin{equation}\label{3.1}
\hat {\cal I} = \frac{e}{2}\frac{d}{dt}(\hat N_r - \hat N_l)
\end{equation}
where in the case of immovable nanoisland only the even mode 1
contributes to the current. In our case both modes 1, 2 are
involved in the tunneling transport due to the term $H_{\rm B}$ in
Eq. (\ref{2.6}). After the Glazman-Raikh rotation the current
operator transforms into
\begin{equation}
\hat {\cal I} = \frac{d}{dt}\hat {\cal Q}_t + \frac{d}{dt}\hat q_t~,
\end{equation}
where
\begin{eqnarray}\label{3.3}
\hat {\cal Q}_t &=& \frac{e}{2} \cos 2\vartheta_t(\hat N_1 - \hat N_2)
\end{eqnarray}
$$
\hat q_t = -\frac{e}{2} \sin 2\vartheta_t\sum_{k\sigma}\left(
  \psi^\dag_{1k\sigma}\psi^{}_{2k\sigma} + \psi^\dag_{2k\sigma}\psi^{}_{1k\sigma}\right)
$$ 
Here the operator $\hat {\cal Q}_t$ controls the time-dependent electron
occupation in the source lead, and the operator $\hat q_t$ is
responsible for all tunneling and cotunneling processes including
admixture of odd components $\psi^{}_{2k\sigma}$ to the
tunneling charge transport induced by the gauge field
$H_B$.\cite{Aono04,KAEW04}

The time-dependent Glazman-Raikh angle defined by (\ref{3.13}) results in adiabatic time dependence of the Breit-Wigner factor
\begin{eqnarray}\label{3.13}
 \sin^2 2\vartheta_t
& =&\frac{4\Gamma_l\Gamma_r}{(\Gamma_l+\Gamma_r)^2} =
\frac{1}{\cosh^2\frac{[y(t)-y_0]}{\lambda}}~,
\end{eqnarray}

 Using the Friedel
-- Langreth sum rule \cite{Langr66}, one may write
\begin{equation}\label{3.4}
N_1 - N_2 = \frac{\delta_t}{\pi}.
\end{equation}
where $\delta_t=\delta_\uparrow+\delta_\downarrow $ is a total time dependent Friedel phase.
At the unitary limit  $\delta_{\uparrow,\downarrow}=\pm\pi/2$.

\color{black}
We are interested in the Kondo effect contribution to the
tunneling current. This contribution is characterized by the spin
dependent scattering phase shift $\delta_\sigma(\varepsilon)$ in
the source lead, which approaches the unitarity limit $\pi/2$ at
  $T$$\to$$ 0$ and $\varepsilon$$\to $$\varepsilon^{}_F$. In the adiabatic limit $\hbar\omega_0 \ll k_B T_K^{min}$ under conditions  
$(k_B T, g\mu_B B, |eV_{\rm bias}|)$ $\ll$ $k_B T_K^{min}$ the phenomenological Fermi liquid Hamiltonian $H_{\rm Noz}$  may be used \cite{Nozieres74} (here $k_B$ is Boltzmann constant, $\mu_B$ is 
Bohr magneton and $g$ is Land\'e factor). 
In this Hamiltonian both scattering and
interaction are scaled by the  time-dependent  Kondo temperature $T_K(t)$ taking minimal value $T_K^{min}$ at maximal distance from the source.

In order to get full tunnel current in adiabatic approximation we  (i) calculate 
a linear response with respect to both bias $eV_{\rm bias} \ll k_B T_K$ and $\hbar d\vartheta_t/dt\leq \hbar\omega_0 \vartheta_{max}\ll k_BT_K$,
 (ii) take into account  cancellations arising due to emergent $SU(2)$ symmetry associated with channels  \cite{com1},\cite{com2a}, (iii) perform averaging with the  adiabatic   Hamiltonian 
 (\ref{2.6}-\ref{2.61}) at zero bias and zero temperature.  The finite temperature and bias effects 
 are accounted by Nozieres method. \cite{Nozieres74}, \cite{com2} 
As a result, the tunnel current $\bar{\cal I}_t =  \bar I_0(t) + \bar I_{\rm int}(t)$ consists of two parts:
the Friedel phase contribution
\begin{equation}\label{3.71}
\bar I_0(t)=\frac{e}{2\pi}\cos
2\vartheta_t\cdot\frac{d\delta_t}{dt}
\end{equation}
and the "ohmic" current \cite{com2b}
\begin{equation}\label{3.72}
\bar I_{\rm int}(t)=\frac{e^2}{\hbar} V_{\rm bias}\frac{d}{dt}\left[\sin 2\vartheta_t
\int_{-\infty}^t dt' \sin 2\vartheta_{t'} \Pi^R_{t-t'}\right]
\end{equation}
where
$$
\Pi^R_t=-\frac{i}{2}\sum_{k\sigma}\sum_{\alpha\neq\gamma}
\big[G^R_{\alpha k\sigma}(t)G^K_{\gamma k\sigma}(-t)+
G^K_{\alpha k\sigma}(t)G^A_{\gamma k\sigma}(-t)\big]
$$
and $G^\Lambda_{\alpha=1,2}$ are bold retarded ($\Lambda=R$), advanced ($\Lambda=A$) and Keldysh 
($\Lambda=K$) Green's function \cite{com2b}.
\color{black}

Let us first rewrite the \color{black} Friedel part of the \color{black} tunneling current (\ref{3.71}) via the
parameters characterizing the Kondo tunneling in the
low-temperature strong coupling limit at $T\ll T_K^{\color{black}min\color{black}}$, where
\begin{equation}\label{3.8}
k_B T_K\color{black}(t)\color{black}=D_0\exp\left[-\frac{\pi E_c}{4(\Gamma_l + \Gamma_r)}\right],
\end{equation}
$D_0$ is the ultraviolet cut-off for the Kondo problem  with a scale of the band energy in
the source, $\Gamma_\alpha =\pi\rho_0 |v_\alpha|^2$, $\rho_0$ is the density of electronic states at the Fermi level $\varepsilon_F$. In our adiabatic regime $T_K$ parametrically
depends on time, following the time dependence of $\Gamma_l(t)$.
The Hamiltonian $H_{\rm Noz}$  
\cite{Nozieres74}, \cite{com2}
establishes the relations between $\delta_{\sigma}$, $B$ and $T_K$ 
near the unitary limit,  such as $\delta_t= 2|e V_{\rm bias}|/(k_B T_K(t))\ll 1$. The magnetic field enters
only into relative Friedel phase $\delta_\uparrow-\delta_\downarrow=\pi - 2g\mu_B B/(k_B T_K(t))$.
\color{black} We neglect the influence of magnetic field on $T_K$, since we work in the limit
$g\mu_B B\ll T_K^{min}$. Alternatively, a non-uniform magnetic field negligible at the dot and gradually  increasing along the cantilever could be assumed in the model.
\color{black}

\begin{center} 
\begin{figure}[t]
\vspace*{-12mm}
\includegraphics[angle=0,width=\columnwidth]{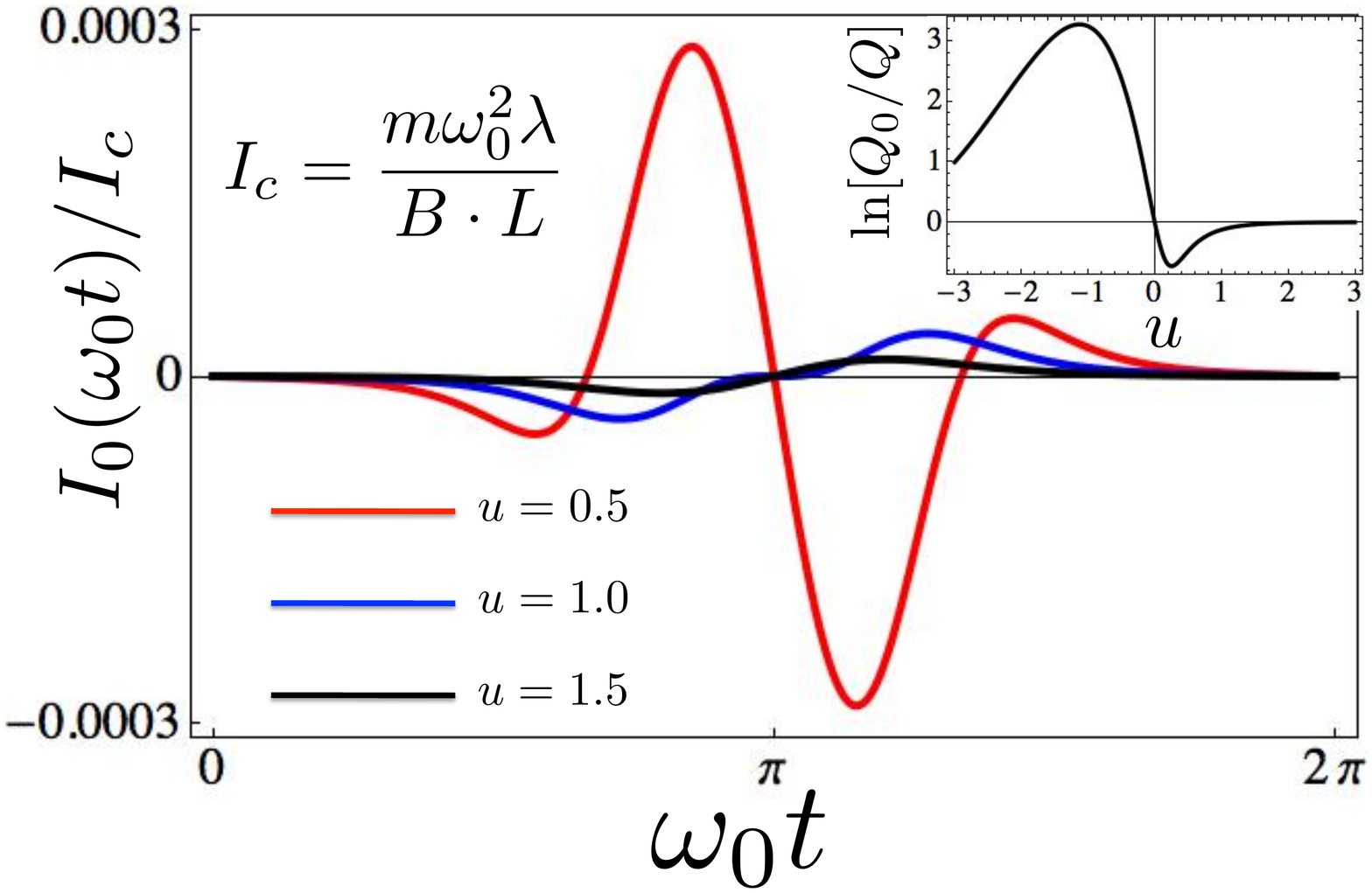}
\vspace*{-13mm}
\caption{(Color online) Time dependence of the current $I_0$ for different values of asymmetry parameter $u=y_0/\lambda$. For all three curves shuttle oscillates with amplitude $y_{max}=\lambda$,  $\hbar \omega_0/(k_B T_K^{\color{black}min\color{black}})=10^{-3}$, $|eV_{\rm bias}|/(k_B T_K^{\color{black}min\color{black}})= g\mu_B B/(k_B T_K^{\color{black}min\color{black}})=0.1$ with $T_K^{(0))}=2 K$, $\lambda/L=10^{-4}$. 
Insert:  $\ln[Q_0/Q]$ as a function of $u$, $Q_0=10^{4}$.}\label{ff.1}
\end{figure}
\end{center}

In the adiabatic limit the Friedel phase $\delta_t$ and Glazman - Raikh angle $\color{black}\vartheta_t\color{black}$ are not independent,
but connected through (\ref{3.13}) and (\ref{3.8})
\begin{eqnarray}
\frac{1}{\delta_t}\frac{d \delta_t}{d t} = \frac{\pi
 E_c}{4\Gamma_0} \sin 2\vartheta_t \frac{d \vartheta_t}{d t} 
\end{eqnarray}
with $\Gamma_0 =\pi\rho_0|v_0|^2\ll E_c$.
Thus, the Friedel contribution to tunnel current can be expressed in terms
of shuttle velocities as follows:
\begin{equation}\label{3.15}
\bar I_0(t) = \frac{\dot y}{\lambda} \frac{e
E_c}{8\Gamma_0}\cdot\frac{e V_{\rm bias}}{k_B T_K(t)}\cdot \frac{
\tanh\left(\frac{y-y_0}{\lambda}\right)}{
\cosh^{2}\left(\frac{y-y_0}{\lambda}\right)}\\
\end{equation}
Here the time dependence of the tunnel current is predetermined
by the time dependence of tunnel integrals for the nanoisland
moving in $y$ direction, i.e. by the function $y(t)$ and its
derivative $\dot{y}$.  Moreover, one can see that even in the case of possible instability
large amplitude oscillations are exponentially suppressed. Typical behaviour of $I_0(t)$ is shown in Fig.\ref{ff.1}.
The non-sinusoidal form of current is associated with time dependence of both tunnel width and Kondo temperature.

The second term $\bar I_{\rm int}(t)$ given by Eq. (\ref{3.72})
leads to "ohmic" contribution to the current with unitary conductance $G_0=e^2/h$ (see discussion below):
\begin{equation}\label{3.16}
\bar I_{\rm int}(t) = G_0 V_{\rm bias} \sin^2 2\vartheta_t\sum_\sigma\sin^2\delta_\sigma
\end{equation}

\vspace*{-10mm}
The force in the r.h.s. of the Newton equation (\ref{2.4}) is a sum of the driving
force $F_0$, the Laplace force $F_L$ and electromotive (emf) force $F_{\rm emf}$:
\begin{equation}\label{4.2}
F(y,t) = F_0(t) +\bar {\cal I}_t\cdot B\cdot L  + F_{\rm emf}
\end{equation}
The emf force can be estimated as $F_{\rm emf}$$\sim$$\dot y$$(B\cdot L)^2$ $G_0 $ \cite{com3}. 
Due to sequential geometry of electric circuit, the current
$\bar {\cal I}_t=\bar I_0(t)+\bar I_{\rm int}(t)$ is the tunneling current defined by (\ref{3.15}) and (\ref{3.16}) .  In the limit of
small bias voltage $|eV_{\rm bias}|\ll k_B T_K^{\color{black}min\color{black}}$, electrons in the source
and the cantilever are supposed to be in adiabatically stationary
state of thermal equilibrium. 
Then the parametrization (\ref{3.15}) is valid and  with accuracy to small 
parameters $O(\left[(eV_{\rm bias} /(k_B T_K^{\color{black}min\color{black}})\right]^2,\left[(g\mu_B B )/(k_B T_K^{\color{black}min\color{black}})\right]^2)$ the Lorentz force may be  written as
\begin{equation}\label{4.5}
F_L = F_{ad}(y(t))-\dot y \frac{dF_{ad}}{dy}\frac{\hbar \pi E_c}{16 \Gamma_0 k_B T^{(0)}_K}
\end{equation}
where  $F_{ad}= 2B\cdot L\cdot  G_0\cdot  V_{\rm bias}\cosh^{-2}\frac{[y(t)-y_0]}{\lambda}$
and $T_K^{(0)}$ is a Kondo temperature at equilibrium position.
 Small correction to the adiabatic Lorentz force  in the (\ref{4.5}) may be 
considered as a first term in the  expansion over a small  non adiabatic 
parameter \color{black} $\omega_0\tau\ll 1$\color{black}, where
$\tau$ is the retardation time associated with inertia of the Kondo cloud. Using 
such interpretation  one gets $\tau$:
\begin{eqnarray}\label{4.6}
\tau= \frac{\hbar \pi E_c}{16 \Gamma_0 k_B T^{(0)}_K}=\color{black}\frac{1}{2}\left|\frac{Q^{-1}(B)-Q^{-1}(-B)}{\omega(B)-\omega(-B)}\right|\color{black}
\end{eqnarray}
\color{black}
where $Q(B)$ and $\omega(B)$ are the quality factor and oscillator's frequency at finite magnetic field $B$
respectively.
\color{black}
Equation Eq.\ref{4.6} allows one  to obtain information about dynamics of the 
Kondo clouds from the analysis of the  experimental investigation of 
the  mechanical vibrations. \color{black} The retardation time  associated with dynamics of Kondo cloud is parametrically large compared with the time of formation of the Kondo cloud $\tau_K=\hbar/T_K$ and can be measured owing to small deviation from adiabaticity.  \cite{com2c} \color{black} 
Also we would like to emphasize a supersensitivity of the quality 
factor to the change of the equilibrium position of the cantilever 
characterizing by the parameter $y_0$.
The plot $\ln[Q_0/Q]$ is presented in insert of  Fig.\ref{ff.1}.
From this plot one can see that both suppression  $Q>Q_0$  and enhancement $Q<Q_0$ of the dissipation of
nanomechanical vibrations (depending on the direction of the magnetic 
field and the equilibrium position of the cantilever ) can be stimulated by  Kondo 
tunneling.  The latter demonstrate potentialities for the Kondo induced 
electromechanical instability which will be a subject for separate 
analysis.

Equations (\ref{3.15},\ref{4.5}) and (\ref{4.6}) represent the central results of the Letter. On the one hand,
we have shown that the electric current associated with the Kondo effect results in magnetic field dependent
$Q$-factor allowing to fine-tune the nano-mechanical resonator. On the other hand, the non-ohmic part of the current provides an information about retardation effects related to the motion of the Kondo cloud.
Thus, the measurement of the Kondo forces in Single Electron Transistor give a complementary to conductance measurements information. 

In conclusion we have shown that the Kondo phenomenon in single electron tunneling
\color{black} gives a very promising and efficient mechanism \color{black} for electromechanical transduction on a nanometer length scale. Measuring of nanomechanical response on Kondo-transport in nanomechanical single-electronic device enables one to study kinetics of formation of Kondo-screening and offers a new approach for studying  nonequilibrium Kondo phenomena.  Kondo effect provides a possibility
for super high tunability of the mechanical dissipation as well as super sensitive
detection of mechanical displacement.

We appreciate illuminating discussions with B.L. Altshuler, P. Brouwer,  J. von Delft, Yu. Galperin, S. Ludwig, F. von Oppen and E. Weig. \color{black} MK acknowledges the hospitality of Institute Henri Poincare (Paris)
at the workshop ''Disordered Quantum Systems'' in April-July 2012,
where part of this work has been done.\color{black}


\begin{thebibliography}{99}
\bibitem{footnote} The form $J_{\alpha\alpha'}$$=$$4v^{*}_\alpha$$v^{}_{\alpha'}$$/$$E_c$ 
of exchange parameter implies a Schieffer-Wolff-like transformation of generic time-dependent Anderson Hamiltonian. The adiabatic part of transformation matrix gives this equation. \cite{KNG} Non-adiabatic corrections 
$\sim$$\hbar$$\omega_0$$/$$E_c$ \cite{KNG,CFK} are neglected.
\bibitem{com1}  One can notice that the time-dependent Berry\color{black}-like \color{black} term $H_B$, the first and the second terms of $\delta H$ in Eq.\ref{2.6} form the $SU(2)$ algebra. The appearance of the Berry phase leads to extra kinematic constraints to be taken into account when the tunnel current is calculated.
\bibitem{com2} The Nozieres \cite{Nozieres74} fixed point Fermi Liquid Hamiltonian 
$H_{\rm Noz}$$=$$\sum_{k\sigma} $$\tilde\xi_k$$\phi^\dagger_{k\sigma}$$\phi_{k\sigma}-$
$\sum_{kk'\sigma} $$\frac{\tilde\xi_k+\tilde\xi_k'}{2\pi \rho_0 k_B T_K}$$\phi^\dagger_{k\sigma}$$\phi_{k'\sigma}+$
$+\frac{1}{\pi \rho^2_0 k_B T_K}$$\sum_{kk'}$$\phi^\dagger_{k\uparrow}\phi_{k'\uparrow}$$
\sum_{kk'}\phi^\dagger_{k\downarrow}\phi_{k'\downarrow}$ 
is written in terms of
$\phi_{k,\sigma}$ fermions with a bandwidth $k_B T_K$ and spectrum $\tilde\xi_k$.
It is used for accounting for inelastic contributions to the transport (see \cite{PuGla04}).
\bibitem{com3} The smallness of electromotive force is guaranteed 
by the upper bound on external magnetic field applied to the cantilever 
$\Phi/\Phi_0\cdot L/\lambda < E_c/\Gamma_0 \cdot 
|eV_{\rm bias}|/(k_B T_K)$ where $\Phi= B\cdot {\cal S}_\lambda$ is a flux through the area ${\cal S}_\lambda\sim\lambda^2$ and $\Phi_0=h/e$ is a flux quantum.  The value of the field corresponding to this bound is roughly estimated as $B_u\sim 10 T$.
\end{thebibliography}
\end{document}